\documentclass[journal]{IEEEtran}
\usepackage{amsmath,amsfonts}
\usepackage{algorithmic}
\usepackage{algorithm}
\usepackage{array}
\usepackage[caption=false,font=normalsize,labelfont=sf,textfont=sf]{subfig}
\usepackage{textcomp}
\usepackage{stfloats}
\usepackage{url}
\usepackage{verbatim}
\usepackage{graphicx}
\usepackage{cite}
\usepackage{hyperref}

\usepackage{hhline}
\usepackage{xcolor}
\usepackage{multirow}
\usepackage{colortbl}
\usepackage{tabularx}

\begin{document}

\title{Tensions between Preference and Performance: Designing for Visual Exploration of Multi-frequency Medical Network Data}

\author{Christian Knoll, Laura Koesten, Isotta Rigoni, Serge Vulliémoz, and Torsten Möller,~\IEEEmembership{Senior Member,~IEEE}
}



\maketitle

\begin{abstract}
The analysis of complex high-dimensional data is a common task in many domains, resulting in bespoke visual exploration tools. Expectations and practices of domain experts as users do not always align with visualization theory.
In this paper, we report on a design study in the medical domain where we developed two high-fidelity prototypes encoding EEG-derived brain network data with different types of visualizations. We evaluate these prototypes regarding effectiveness, efficiency, and preference with two groups: participants with domain knowledge (domain experts in medical research) and those without domain knowledge, both groups having little or no visualization experience.
A requirement analysis and study of low-fidelity prototypes revealed a strong preference for a novel and aesthetically pleasing visualization design, as opposed to a design that is considered more optimal based on visualization theory. Our study highlights the pros and cons of both approaches, discussing trade-offs between task-specific measurements and subjective preference. While the aesthetically pleasing and novel low-fidelity prototype was favored, the results of our evaluation show that, in most cases, this was not reflected in participants’ performance or subjective preference for the high-fidelity prototypes.
\end{abstract}

\begin{IEEEkeywords}
Design study, EEG data, network analysis
\end{IEEEkeywords}

\section{Motivation}
\label{sec:introduction}
\IEEEPARstart{T}{he} visualization community developed many helpful design guidelines to greatly narrow the possible approaches for developing visual analysis interfaces, from facetting (and the general approach of interactive multi-view dashboards) to Shneiderman's Mantra~\cite{shneiderman1996}. However, all of these guidelines optimize the effectiveness of such visual analysis systems. When a new goal, such as aesthetics and engagement, is being formulated, it becomes again a challenge to narrow the design space properly. This paper reports on a design study to help clinical researchers explore and understand complex high-dimensional data from electroencephalography (EEG) and epilepsy patients. Specifically, they compute and analyze the specific network of connections between brain regions across different EEG frequencies (frequency bands) where simple visualization approaches have not helped the analysis. However, our domain experts very much favor novel and engaging interfaces.

One of the non-trivial challenges of this process is to develop tools that (a) embed properly in an existing workflow and (b) respect certain expectations of the domain experts. From the beginning, our domain experts communicated to us their expectations of engaging visual encodings that would be novel within their particular domain. Due to their aesthetics, they favored particular visual encodings (circular design) early on in the prototyping process. Thus, we were interested in the challenge of developing the best prototype under the constraint of an engaging circular layout and comparing it to the best prototype based on alternative designs, following a traditional functional design approach. We are unaware of other design studies that have dealt explicitly with these competing design goals.  

Because of our domain experts' bias toward novelty and aesthetics, it was important for us to test and collect feedback on our prototypes from people outside of the specific domain as well. While there are many definitions for domain novices~\cite{burns2023}, we refer to them as lay people without domain expertise having little or no visualization experience. The drawback was to find a way to give them tasks to be done with our prototypes that they could understand without domain knowledge. The advantage would be that they have likely not dealt with similar visualization problems before and might be less biased in either direction (aesthetics vs. functionality). At the same time, however, it was essential for our evaluation to not just measure overall usability (and functionality) as in a typical design study but to measure perceived aesthetics and engagement specifically.

We developed two high-fidelity prototypes encoding multi-frequency EEG data. We evaluate these prototypes regarding effectiveness, efficiency, and subjective preference with participants with domain knowledge (domain experts) and without domain knowledge (lay people), both groups having little or no visualization experience.
The primary contributions of this paper include:
\begin{itemize}
    \item a characterization of the domain problem and tasks through an in-depth requirement analysis with diverse medical researchers working on multi-frequency network data (\autoref{sec:methodology}).
    \item two high-fidelity prototypes with different visual encodings both optimized for functionality. However, one was additionally constrained by an aesthetically pleasing visual encoding (\autoref{sec:designdecisions}).
    \item a user study comparing four domain experts' and twelve lay peoples' subjective preferences and objective task performance (\autoref{sec:evaluation}).
\end{itemize}

\section{Domain Background}
\label{sec:domainbackground}
Recently, EEG-based network analyses have been increasingly applied to investigate brain networks in physiological conditions or brain disorders such as epilepsy~\cite{barollo2022,carboni2020,royer2022}. Promising results have been found for both the diagnosis and prognosis of epilepsy patients. At the core of the analysis are novel approaches for estimating and characterizing the brain network of a particular patient. These networks give insight into the brain's function (or dysfunction). However, insufficient validation of these brain networks limits the applicability of analysis techniques for clinical practice~\cite{vanmierlo2019}.

Among issues linked to the multiple ways of estimating connectivity and characterizing functional graphs, the visualization of graphs and measures within them is complex. Simplified visualizations, as used commonly in the domain, further hinder the interpretation of these networks. Typically, for brain networks derived from high-density EEG (HD-EEG), the number of (cortical) regions investigated ranges from 64~\cite{desikan2006} to 72 to even more regions, depending on specific research needs~\cite{hagmann2008,rigoni2023}. Furthermore, EEG data is usually analyzed in several frequency bands to reflect specific activity patterns related to cognition and pathological alterations~\cite{cohen2017}. Graph measures can be computed at the global level (e.g., to describe the global integration or segregation of the graph), at the hemispheric level, or even at the nodal one~\cite{rubinov2010}. Significant differences between nodal measures (e.g., across patients, vigilance state, before and after medication, etc.) are much harder to visualize due to the high number of nodes and frequency bands in the brain network, not to mention the temporal dimension in some studies.

Therefore, we developed a tool that provides a descriptive visual view of nodal graph measures (e.g., the clustering coefficient) and visualizes them across frequency bands or as part of different subnetworks. The goal of this tool is, on the one hand, to enable researchers to gain a descriptive view of the network before statistical testing. It should allow them to formulate hypotheses based, for example, on perceptual similarities across frequency bands for a specific cluster of nodal regions of interest (ROI). This would enable exploratory visual analyses of the data without investigating each brain region separately, leading to more powerful targeted analyses. It eliminates the need to adjust significance levels for multiple testing~\cite{bender2001}. 
On the other hand, it could, for example, enable clinicians to explore different network patterns arising from removing some nodes to simulate a lesion or a surgical procedure, facilitating the incorporation of research data into clinical discussions and promoting interactive multidisciplinary meetings.

\section{Related Work}
\label{sec:relatedwork}
In this section, we review existing literature on the effectiveness of visualization encodings, different factors influencing a visualization design, and visual analysis tools for EEG data.

\subsection{Visualization Marks and Channels}
To create effective visualizations, designers must choose the optimal encoding for their data. Visual encodings consist of marks and channels. Marks are graphical elements in visual encodings, and their appearance is controlled through channels~\cite{munzner2014}. The effectiveness of a channel depends on the data type. We primarily use the length of a bar and color saturation as channels in our prototypes---the former being the more effective encoding for our data based on visualization literature~\cite{munzner2014,cleveland1984,mackinlay1986,heer2010}. These studies show that spatial channels most effectively encode ordered data. A recent survey by Quadri and Rosen~\cite{quadri2022} provides an overview of existing perception-focused visualization studies, categorizing them by the task taxonomy of Amar et al.~\cite{amar2005}.

\noindent \textbf{Bar length as a channel.} There are several studies investigating the perception of bar charts. The visualization tasks discussed in our work commonly relate to visual comparison, e.g., of bars across five frequency bands and up to 72 ROIs, as detailed in \autoref{sec:requirement}.
For instance, the visual comparison of bars has been researched by Talbot et al.~\cite{talbot2014}, who found that comparing non-adjacent bars is difficult, especially for short bars. A study by Nothelfer and Franconeri~\cite{nothelfer2020} looked at how relations between bars are perceived and found that data deltas were better processed visually when explicitly coded. This is also suggested by Srinivasan et al., who evaluated different bar chart variants for visual comparison and found that adding ``difference overlays facilitate a wider range of comparison tasks''~\cite[p. 1]{srinivasan2018}. While most of these studies focus on comparing a pair or a small number of bars, one of our prototypes displays up to 72 bars in each of the five bar charts.
Gramazio et al.~\cite{gramazio2014} showed that size, quantity, and grouping of marks influence user performance and found that search tasks can be solved faster when marks are spatially grouped (such as in bar charts) rather than randomly arranged. Additionally, linear arrangements of marks are usually more effective~\cite{waldner2020} and efficient~\cite{diehl2010} than radial arrangements -- a comparison we also touch upon in our prototypes.

\noindent \textbf{Color as a channel.} Using color in visualizations has also been broadly researched. Szafir~\cite{szafir2018} conducted experiments to measure the perception of color difference in visualizations. She found that ``elongated marks provide significantly greater discriminability for encoding designers''~\cite[p. 392]{szafir2018}. Color often encodes continuous quantitative data (e.g., on maps), such as the EEG data we used in our design study. While rainbow color maps are usually considered a poor choice~\cite{borland2007,reda2023,golbiowska2022}, Khairi et al.~\cite{khairi2018} observed that rainbow schemes can be effective for quantity estimation, while divergent colormaps facilitate the perception of high-frequency patterns. We used single-hue sequential colormaps in our prototypes informed by the findings of Karim et al.~\cite{karim2019} where they compared different single- and multi-hue colormaps in network visualization. They found that participants in their study completed tasks significantly faster with a blue single-hue sequential colormap. There are also tools~\cite{gramazio2017,harrower2003} that enable finding effective color schemes for a visualization, which we used to find effective colormaps.

\subsection{Influencing Factors for Visualization Design}
Visualization design decisions are influenced by the appropriate visual encoding and several other factors (e.g., aesthetics, familiarity) that need to be considered when creating effective visualizations. Tory and Möller~\cite{tory2004} argue that human factors (e.g., visual acuity, culture, previous experience) must be considered in designing and evaluating visual analysis tools to add to the systems' usefulness for future users. While human-centered design approaches for visualization systems usually involve a requirement analysis to characterize the domain problem and prospective users, visualization designers often focus primarily on task-based objectives (e.g., conveying facts or insights), and only recently, there has also been a shift towards affective objectives by the academic visualization community. Lee-Robbins and Adar also emphasize paying attention to affective intents ``that seek to influence or leverage the audience’s opinions, attitudes, or values''~\cite[p. 1]{leerobbins2023}. Pandey et al.~\cite{pandey2014}, who examined persuasion of data visualizations, also reflected on affective intent. They found that people's attitudes about a topic influence the persuasive power of data visualization. In our study, we were especially interested in the influence of aesthetics of visualizations~\cite{lau2007,filonik2009,vandemoere2011}. He et al.~\cite{he2023} introduced a novel scale called \textit{BeauVis} for measuring and comparing the aesthetic pleasure of visual representations. Another important factor to consider is the familiarity of future tool users with the visual medium~\cite{dasgupta2017,dasgupta2018}.

\subsection{Visualization Tools for EEG Data}
Visualization is widely used for displaying and analyzing multi-channel EEG data. While all of the following papers use visualization to facilitate EEG analysis, to the best of our knowledge, there are no bespoke tools tailored to domain experts' needs for exploratory EEG data analysis. 
Slaybeck et al.~\cite{slayback2018} report on novel visualization methods for EEG signals using a virtual reality system and a physical head model.
A parallel coordinate method is proposed and evaluated by ten Caat et al.~\cite{caat2007}. Since visualizing connectivity in EEG data as a graph layout can result in visual clutter, ten Caat et al.~\cite{tenCaat2008} propose a graph layout based on functional units (data-driven region of interest). We alleviated the issue of visual connectivity clutter by allowing users to filter the connections.
Wulandari et al.~\cite{wulandari2018} developed a system to visualize the brain wave signals of epilepsy patients. Fang et al.~\cite{fang2019} created a visualization dashboard for comparative analysis. There is also literature on using 3D representations for analyzing EEG data~\cite{gavrilescu2015,mullen2013,christopher2013}. FCLAB~\cite{pezoulas2018} is a plugin-based environment that visualizes brain functional connectivity networks with local and global measures. While this system enables network analysis for each frequency band at the electrode level, we focus on source-reconstructed EEG at the ROI level. Additionally, our prototypes allow for comparing metrics across frequency bands and for user-defined subnetworks.

\section{Design Process}
\label{sec:methodology}
We realized a three-step design process that follows the design study methodology framework~\cite{sedlmair2012}. In the first phase, we interacted with medical domain experts from diverse fields (medical doctors, biomedical engineers, and computer scientists) to identify the challenges they face in their data analysis and to gather ideas about how data visualization can help facilitate this process. We conducted an in-depth requirement analysis with domain experts on-site in the second phase. This requirement analysis was the basis for three digital pen-and-paper prototypes. We used the collected knowledge and iterative feedback cycles in the third phase to implement two interactive high-fidelity prototypes.

\subsection{Ideation}
To generate and explore promising ideas that support the analysis of our collaborators through visualization, we conducted a creative visualization-opportunities (CVO) workshop~\cite{kerzner2019} early on in the project. Nine medical domain experts (five male, four female), including clinicians and neuroscientists, participated in a full-day workshop to explore data visualization opportunities. The workshop was composed of four guided activities: After a short introduction of the topic and participants at the beginning of the workshop, the aim of \textit{wishful thinking} was to generate a broad spectrum of ideas mainly focusing on the domain experts' information needs and the potentially related data visualization tasks.
We gathered the answers on sticky notes and clustered them into common themes. The results showed the participants' interest in different aspects of EEG data, focusing on, among others, the diagnosis, prediction, and nature of epilepsy in patients.

In the second part of the CVO workshop, the participants were shown existing medical or related data visualizations to inspire their thinking and prepare them for the last activity. The workshop concluded with \textit{video prototyping}~\cite{knoll2020}, where participants were asked to develop a low-fidelity prototype in small groups for one of the themes that were found earlier. The CVO workshop helped characterize the domain experts' analysis challenges, gather ideas on how visualization could facilitate the analysis process, and produce the first prototypes that showcase how a visual analysis interface could look. We also identified collaborators who were excited to work with us on this project. 

\subsection{Requirement Analysis}
\label{sec:requirement}
In the second phase, we analyzed the artifacts produced in the initial workshop and used this knowledge to plan an in-depth requirement analysis with our collaborators. We spent one week working closely with the domain experts at their hospital to define the requirements for a visual analysis tool to facilitate their analysis of multi-frequency EEG data. Interviews with four domain experts (medical doctors, neuroscientists, and engineers) about their research, focusing specifically on the data they analyze and the context of data use, gave us an in-depth understanding of their analysis processes.

\begin{figure}[tb]
    \centering
    \includegraphics[width=0.35\textwidth]{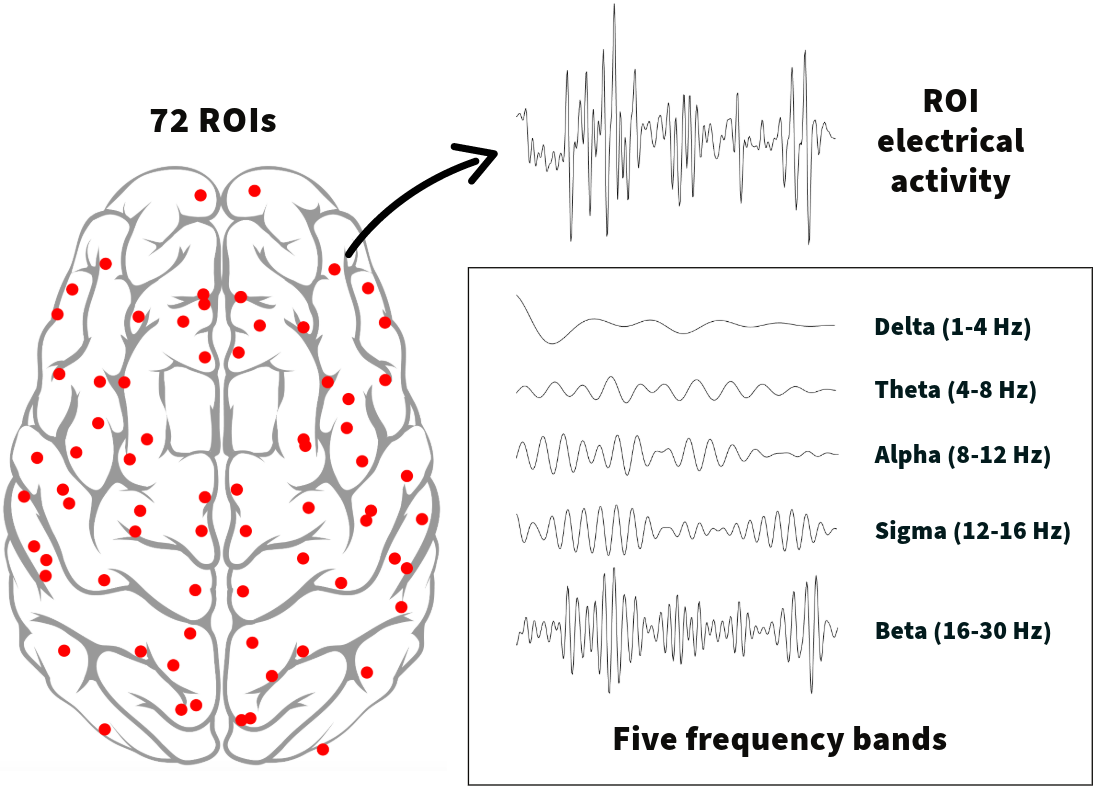}
    \caption{Schematic representation of the EEG data. The grey matter (outer layer of the brain) was parceled into 72 ROIs (red dots), and each ROI's electrical activity was reconstructed from the EEG signal. The frequency content of the ROI signal is displayed for each frequency band in the box. Once the connectivity between each ROI is estimated, a brain network (ROI x ROI) is built for each frequency band. Then, nodal network metrics – such as the clustering coefficient – are computed for each ROI.}
    \label{fig:eegdata}
\end{figure}

\noindent \textbf{Data.} The domain experts mainly operate in a clinical setting and look at multidimensional EEG data (see \autoref{fig:eegdata}) of epilepsy patients over time. EEG is a non-invasive technique used to measure the electrical activity in the brain using electrodes placed on the scalp. The data consists of a time series of electrical voltages reconstructed for different areas of the brain (regions of interest, ROI). By quantifying the level of similarity between time series belonging to different ROIs, the domain experts build a connectivity matrix that describes statistical dependencies between different brain regions. As different frequency components have cognitive and pathological correlates, domain experts analyze the brain network across frequency bands. In other words, the final EEG connectivity data is a multidimensional matrix (ROI x ROI x frequency), where ROI's are the nodes and their relations are the network's edges. Finally, graph metrics are computed to characterize the brain graph at different frequencies, for example, by describing its segregation. More specifically, metrics such as the clustering coefficient (CC) are computed for each ROI and frequency band and describe the local level of clusterization between neighboring nodes, reflecting functional specialization. More details about EEG data collection, preprocessing, and connectivity analyses are provided as supplemental material. We used the knowledge gathered in the first part of the week to develop three low-fidelity prototypes in a digital pen-and-paper setting, which will be described in detail in \autoref{sec:lowfi}.

\noindent \textbf{Tasks.} The domain experts focus their analyses on the whole brain network and the connections between specific ROI, and they extract global and nodal graph measures to describe these interactions. Whether this is done in broadband or for different frequency bands depends on the research focus. As mentioned, our data set consisted of five connectivity matrices, one for each frequency band (delta, theta, alpha, sigma, and beta), representing a typical data set for domain experts. The interview results show that the domain experts are especially interested in exploring nodal graph measures across frequency bands, as global measures are usually easier to display (typically in a boxplot). To summarize, the domain experts need a tool for the visual exploration of EEG data that facilitates the following tasks:
\begin{itemize}
    \item \textit{Overview:} getting an overview of ROIs and their connectivity for different frequency bands
    \item \textit{Cluster:} grouping of ROIs in subnetworks
    \item \textit{Locate:} locating ROIs and connections with high activity
    \item \textit{Compare:} comparing ROIs and connections within and across frequency bands
\end{itemize}

\subsection{Implementation}
The last phase of our process focused on the prototype transition from low-fidelity to high-fidelity. Based on the results of the evaluation of the low-fidelity prototypes through interviews with the domain experts, we decided on two high-fidelity designs (see \autoref{sec:hifi}) for further implementation. Two interactive visualization dashboards were developed while conducting regular feedback cycles with the domain experts. Each dashboard consists of a JavaScript frontend that uses D3.js~\cite{bostock2011} for visualizing the data.

\section{Design Decisions}
\label{sec:designdecisions}
In this section, we will detail our design decisions during prototyping. First, we discuss our low-fidelity designs, where we created multiple digital pen-and-paper prototypes in parallel to explore different design options. Next, we elaborate on interviews that we conducted with our domain experts to assess these low-fidelity prototypes. We conclude this section by reporting on how we arrived at our high-fidelity prototypes.

\begin{figure*}%
    \centering
    \subfloat[Lo-fi prototype 1]{
    \includegraphics[width=0.32\textwidth]{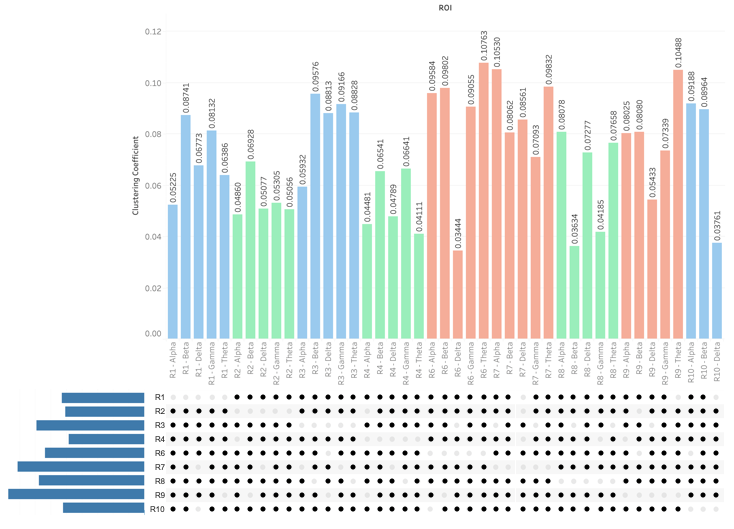}
    }\qquad
    \subfloat[Lo-fi prototype 2]{
    \includegraphics[width=0.30\textwidth]{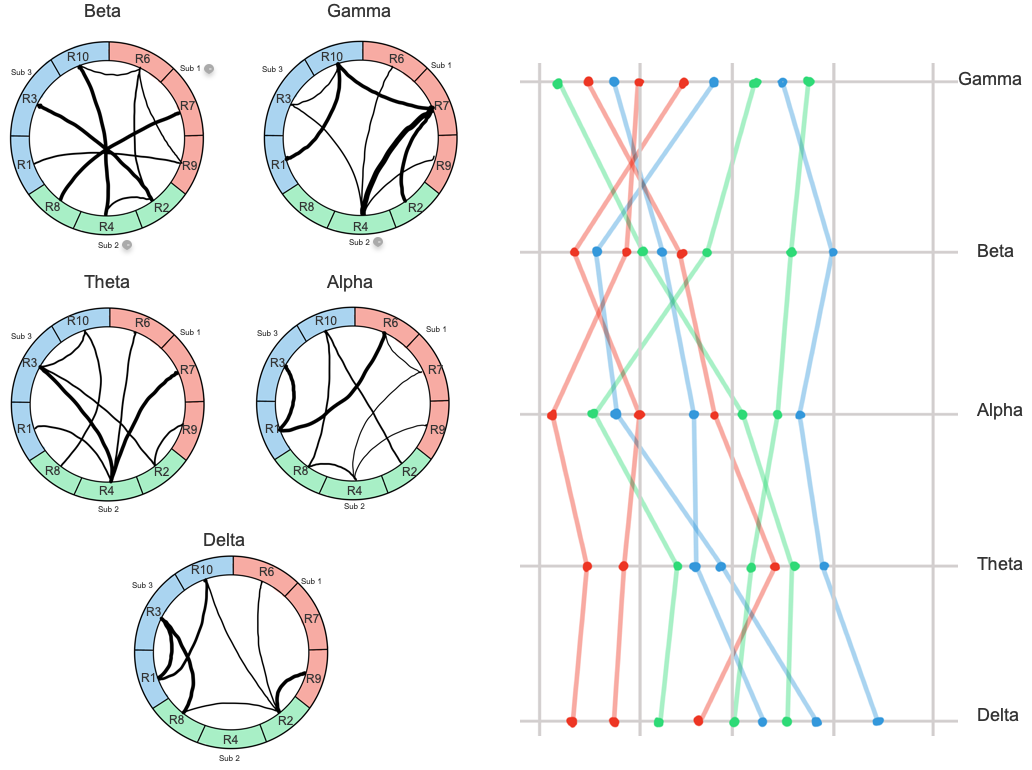}
    }\qquad
    \subfloat[Lo-fi prototype 3]{
    \includegraphics[width=0.22\textwidth]{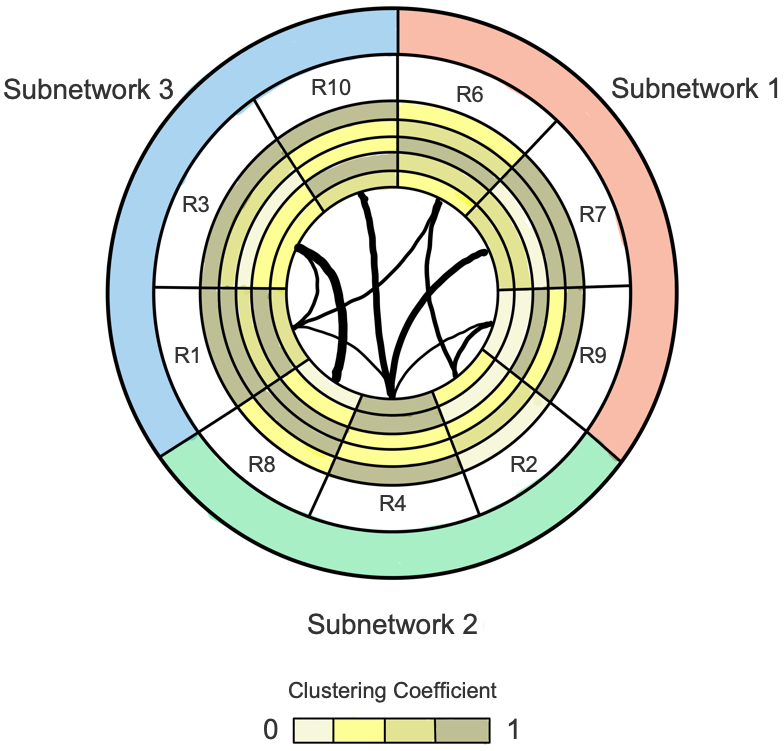}
    }
    \caption{The three low-fidelity prototypes: (a) using bar charts (metric) and a dot plot (connectivity), (b) using parallel coordinates (metric) and doughnut charts (connectivity), and (c) using a layered doughnut chart (metric and connectivity). \textit{Remark: user interface elements (e.g., menu for selecting ROIs) have been excluded to focus on the visual encoding of the data.}}
    \label{low-fi}
\end{figure*}

\subsection{Low-fidelity Designs}
\label{sec:lowfi}
From the in-depth requirement analysis conducted on-site at our collaborators' hospital, we identified the researchers' need for a tool that facilitates the visual investigation of EEG-derived brain network data. While they primarily use mathematical analysis to explore the data, inspection of results through visualization occurs only as a final step of their analysis due to difficulties linked to visualizing complex multidimensional data. Finding an intuitive visual encoding of the data is necessary for several reasons, including but not limited to summarizing, validating, and communicating their findings. Based on this knowledge, we developed three low-fidelity prototypes (see \autoref{low-fi}) that use different visual representations to facilitate analysis. Detailed information about the low-fidelity prototypes can be found in the supplemental material.
Four domain experts were interviewed to evaluate the three low-fidelity prototypes. A ranking of the designs revealed a preference for low-fidelity prototype 3 (see Figure \ref{low-fi}c) as it was ranked first in three of the four interviews. The participants found this design the most intriguing and innovative since they had never worked with this type of aggregated visualization before. They especially liked the idea of layering multiple frequency bands in segments on top of each other in a single circular layout, allowing for comparison across the frequencies. However, one participant found the design less intuitive than the other two, and the limited space for the connections inside the circle was pointed out as a weakness by another participant.

Low-fidelity prototype 1 (see Figure \ref{low-fi}a)  was ranked first by one participant and ranked in second place (out of three) by two other participants. Participants described the bar charts as intuitive, and one added that they allow for identifying trends. Grouping the bars by frequency band was identified as helpful by one of the participants since they can compare values within a frequency and between frequencies. The primary issue that three of the four participants identified was related to the tool's scalability. The number of ROIs is usually relatively high (from 72 ROIs and above), so showing all the regions in the bar chart might not effectively scale. As a follow-up, we showed them an equivalent stacked bar chart (the five frequencies stacked for each region) to understand whether this would help the scalability of the visualization. The feedback suggested that this might be helpful for comparisons between the frequency bands. Still, three of the four participants did not find the stacked bar chart helpful since comparing values for specific frequencies across different regions was harder. All participants liked the dot plot encoding the connections between the regions. However, most reported that they would expect a heatmap to be more informative and intuitive since they are used to working with heatmaps. The horizontal bar chart left to the dot plot was introduced as an optional feature for users to display data attributes of their choice. The participants found this option intriguing but could not imagine a usage scenario. 

The least preferred low-fidelity prototype was design 2 (see Figure \ref{low-fi}b). Two participants pointed out that circular plots are commonly used for plotting connections between entities in the medical research literature, giving them a sense of familiarity. Most participants also liked the idea of having individual circles for each frequency. However, one participant raised concerns about edge clutter with increasing connections and suggested letting users switch between a node-link and an adjacency matrix representation. The parallel coordinates plot did not add any value for three of the four participants since they reported it to be less intuitive than the bar charts. The other participant, however, found it helpful to follow the value changes across frequencies.

\begin{figure*}
    \centering
    \subfloat[Prototype A]{
        \includegraphics[width=0.50\textwidth]{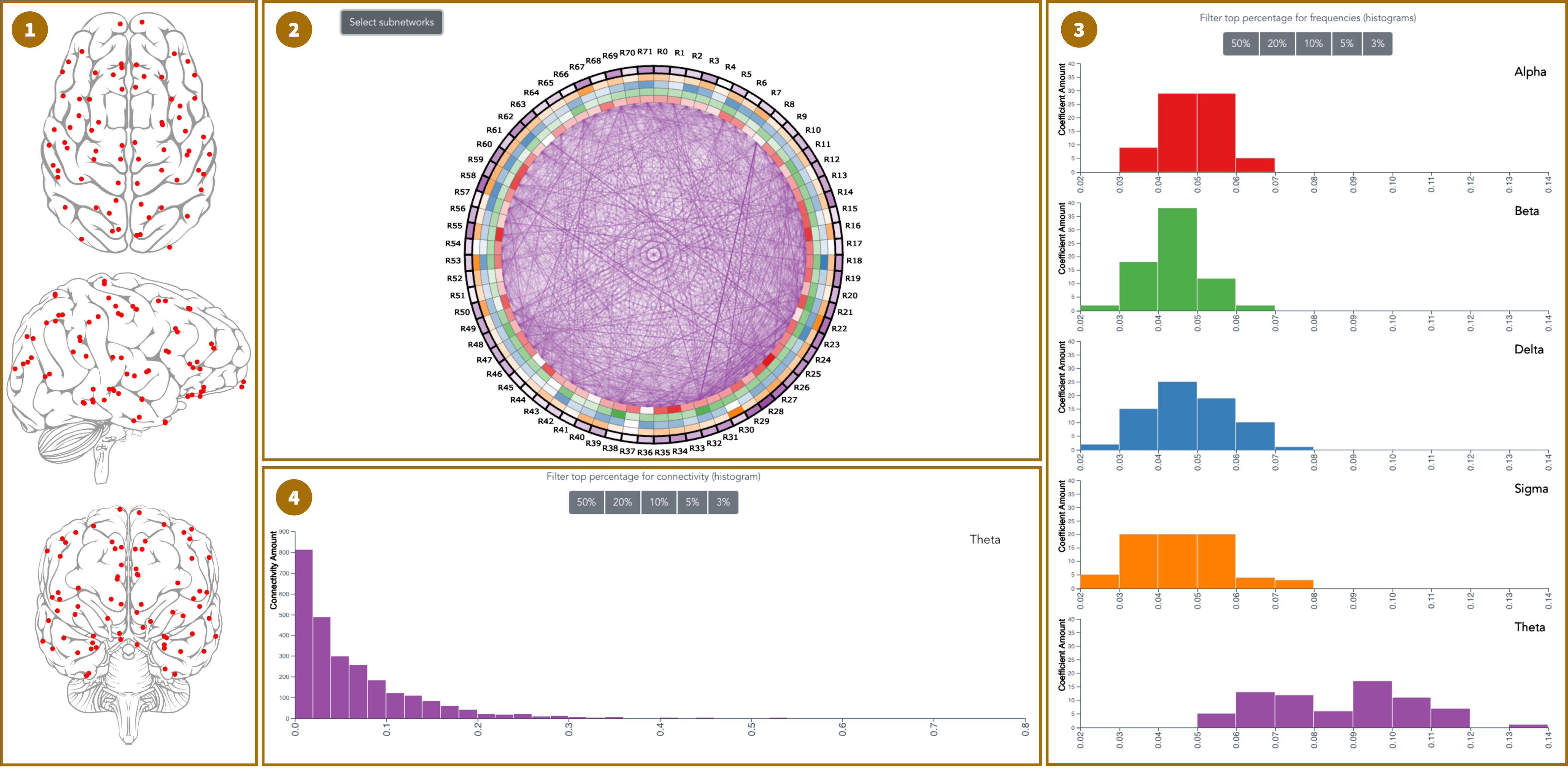}
    }   ~
    \subfloat[Prototype B]{
        \includegraphics[width=0.473\textwidth]{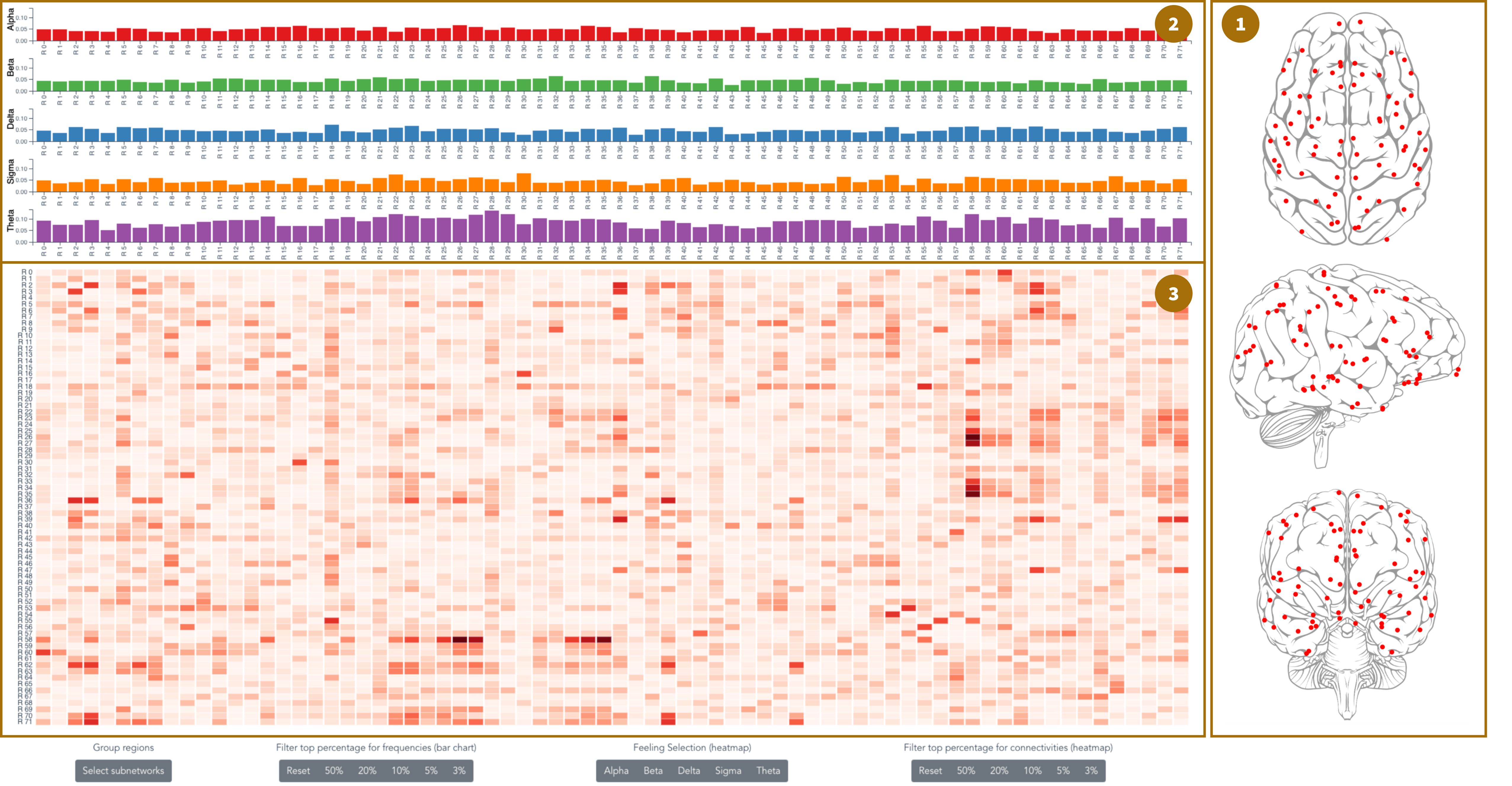}
    }   ~
    \caption{High-fidelity prototypes with all 72 ROIs. Prototype A (based on low-fidelity prototype 3) shows a layered doughnut chart (metric and connectivity), histograms (metric and connectivity), and a brain plot with 72 ROIs. Prototype B (based on low-fidelity prototype 1) shows bar charts (metric), a heatmap (connectivity), and a brain plot with 72 ROIs.}
    \label{fig:hifiprototypes}
\end{figure*}

\subsection{High-fidelity Designs}
\label{sec:hifi}
The interview study showed that low-fidelity prototype 3 was the most favored of our three initial designs. As pointed out by several participants, combining different data features into a single chart was unconventional and new to them. Even though prototype 1 received good feedback, participants seemed to prefer the originality of prototype 3 over conventional chart types. Since their feedback was based on looking at the digital low-fidelity prototypes without any interactivity, we implemented prototype 1 and prototype 3 as testbeds for further evaluation. Prototype 2 was disregarded from further implementation as the interviews showed that the parallel coordinates plot did not add much value to the analysis of our domain experts. We iteratively improved the implemented prototypes with feedback from the collaborating domain experts throughout the implementation phase.\\\\
\textbf{Prototype A} \textit{(layered doughnut chart)} (see Figure \ref{fig:hifiprototypes}a) was implemented based on low-fidelity prototype 3. The final testbed consists of four main views. The \textbf{brain view (1)} \textit{(Overview, Locate)} shows a schematic representation of a patient's brain from three perspectives (from top to bottom): superior view, lateral view, and posterior view. The red dots represent the 72 nodal regions. The \textbf{ring view (2)} \textit{(Overview, Locate, Compare)} contains two parts. The five rings show the values of a metric (in this case, the clustering coefficient) encoded with color. Each ring segment encodes the value for an ROI in different frequencies, and each of the five frequencies is depicted through a color (red, green, blue, orange, and purple). The color saturation of each segment encodes the metric value: brighter colors indicate lower values than darker shades. This allows users to get a first overview of the value distribution. Hovering over a segment, a tooltip details the exact value.

The area within the five rings encodes the connectivity between the different regions. Since the connectivity changes for every frequency, the user can select the frequency by clicking anywhere in the corresponding frequency ring. We used the color saturation of the lines to encode the connection strength to alleviate the problem that there might be up to 72x72 connections where it would be impossible to see individual lines. This should also help users visually identify potential connectivity patterns more easily.

\begin{figure}[tb]
    \centering
    \includegraphics[width=0.40\textwidth]{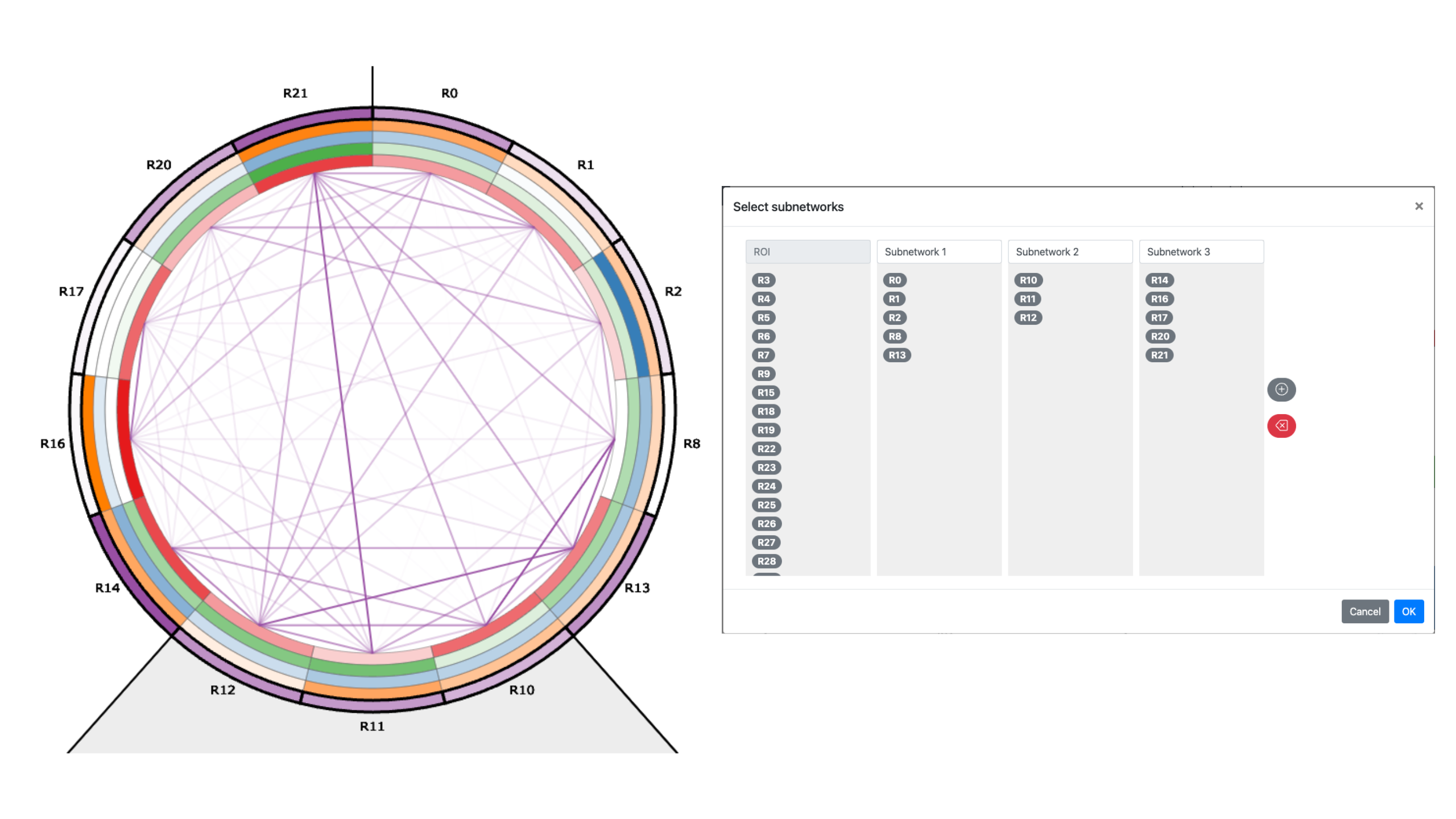}
    \caption{High-fidelity prototype A with three subnetworks containing five, three, and five ROIs, respectively.}
    \label{fig:ringsub}
\end{figure}

The button in the upper left corner opens a popup window that lets the user select subnetworks \textit{(Cluster)}. A user might want to focus only on specific groups of ROIs (here called subnetworks) rather than looking at the global network. \autoref{fig:ringsub} shows three subnetworks containing five, three, and five ROIs, respectively. The lines outside the circle visually indicate the subnetworks.

The \textbf{metric histogram view (3)} \textit{(Overview, Locate, Compare)} on the right shows the distribution of the metric values in each frequency (ring segments). The histograms have the same scale on the y-axis, which enables visual comparison across the different frequencies. Users can select ranges in each histogram for filtering. When a range is selected, only the segments in the ring that are within this range stay colored. The other ones are visually moved into the background by gray grid lines. Each frequency can be filtered individually. In the interview study, we learned that the domain experts are often interested in the top [X]\% of values. Hence, we implemented the buttons above the histograms that allow for a fast selection of those values.

The last \textbf{connectivity histogram view (4)} \textit{(Overview, Locate, Compare)} at the bottom in the middle of the interface shows the distribution of the connection strengths (lines within the circle). The user can again use a range filter, and all connections not contained within this range are not displayed. We also implemented brushing and linking throughout the dashboard to identify the selected ROI(s) in all views.
\\\\
\textbf{Prototype B} \textit{(bar charts and heatmap)} (see Figure \ref{fig:hifiprototypes}b) represents the design from low-fidelity prototype 1 and consists of three main views. The \textbf{brain view (1)} \textit{(Overview, Locate)} is the same as in prototype A and represents the patient's brain with 72 ROIs. On top of the interface, five bar charts show the metric value of every ROI in each frequency. This \textbf{bar view (2)} \textit{(Overview, Locate, Compare)} allows users to compare the values through the length of the bar rather than by color saturation (as in prototype A). The bar charts have the same scale on the y-axis to visually compare the different bar charts. A tooltip showing the exact value appears while hovering over a bar.

The last part of this interface is the \textbf{heatmap view (3)} \textit{(Overview, Locate, Compare)}. This view displays the connectivity between the ROIs for a selected frequency, and each column (ROI) is vertically aligned with the corresponding bars. Connectivity values are encoded through color saturation: brighter colors indicate lower values than darker shades. Hovering a cell allows the user to see the exact connectivity value. Additionally, we support navigation in the heatmap by highlighting the cell's row and column as well as the bars in the \textit{bar view (2)}. On the bottom of this interface, the user has the same options as in prototype A: selecting subnetworks, selecting a frequency for the connectivity to be displayed in the heatmap, and filtering the top percent of values for the metric and connectivity.

\section{Evaluation}
\label{sec:evaluation}
We conducted a study where we used three methods to evaluate the effectiveness, efficiency, and subjective preference of the two high-fidelity prototypes: 
\begin{itemize}
    \item a task-based within-subject study where participants solve four tasks on each prototype while thinking out loud,
    \item a semi-structured interview after completion of the tasks, and
    \item a questionnaire aiming at usability and aesthetics.
\end{itemize}

\begin{table*}
    \centering
    \caption{Demographics of the participants, including their overall prototype preference.}
    \bgroup
    \renewcommand{\arraystretch}{1.1}
    \begin{tabular}{ |c|c|c|c|c|c|c|c| }
      \hline
      & \textbf{ID} & \textbf{Gender} & \textbf{Age} & \textbf{Highest Degree} & \textbf{Job} & \textbf{Domain} & \textbf{Overall Preference}\\
      \hline
      \multirow{12}{*}{\rotatebox[origin=c]{90}{Lay people}} & P$_{1}$ & Male & 27 & Master & PhD student & Computer science & A\\
        & P$_{2}$ & Male & 29 & Bachelor & Full-time employment & Computer science & B\\
        & P$_{3}$ & Male & 20 & Apprenticeship & Full-time employment & Relocation management & A\\
        & P$_{4}$ & Male & 28 & Master & Full-time employment & Software development & B\\
        & P$_{5}$ & Male & 28 & Master & Full-time employment & Civil engineering & B\\
        & P$_{6}$ & Female & 27 & Master & Full-time employment & Computer science & B\\
        & P$_{7}$ & Male & 29 & Bachelor & Part-time employment & Computer science & A\\
        & P$_{8}$ & Female & 29 & Master & PhD student & Computer science & A\\
        & P$_{9}$ & Male & 28 & Master & Part-time employment & Computer science & B\\
        & P$_{10}$ & Female & 25 & High school & Bachelor student & Statistics & B\\
        & P$_{11}$ & Female & 51 & High school & Full-time employment & Social work & B\\
        & P$_{12}$ & Female & 55 & High school & Full-time employment & Public service & B\\
      \hhline{========}
      \multirow{4}{*}{\rotatebox[origin=c]{90}{Experts}} & E$_{1}$ & Male & 35 & PhD & Postdoc researcher & Clinical neuroscience & B\\
        & E$_{2}$ & Male & 34 & PhD & Postdoc researcher & Neuroscience & A\\
        & E$_{3}$ & Male & 27 & Master & PhD student & Biomedical engineering & B\\
        & E$_{4}$ & Male & 31 & PhD & Postdoc researcher & Clinical neuroscience & B \\
       \hline
    \end{tabular}
    \egroup
    \label{tab:participants}
\end{table*}

\subsection{Participants}
We targeted two participant groups to evaluate the two prototypes: participants with domain knowledge in medical research (domain experts) and without domain knowledge (lay people), both having little or no visualization experience (see \autoref{tab:participants}). We recruited four participants between 27 and 35 years old and twelve participants between 20 and 55 for the expert and lay groups, respectively, through purposive sampling. While none of them have advanced visualization experience, they are familiar with basic chart types. Despite our effort to aim for gender-balanced recruitment in both participant groups for the study, we, unfortunately, had no female participants in the domain expert group. This is because this specific data set is currently mainly investigated by male researchers in the research lab which we collaborated with. The study structure was mostly the same for both groups, with a minor difference in how the data was explained (see \autoref{taskstudy}). The tasks used in the lay-participant group did not require domain knowledge and focused solely on the visualizations. We believe the lay participants' results strengthen the user study as they can be used to compare and validate the domain experts' results. We conducted individual sessions with nine participants at our research lab, while the others were conducted through remote sessions. All participants gave written consent to participate in the user study and none of them reported any color vision deficiencies.

\subsection{Task-based Within-subject Study} \label{taskstudy}
Our two interfaces use the Vue.js framework combined with D3.js \cite{bostock2011} to draw the charts. They were optimized to be displayed in the Google Chrome web browser on a 27-inch monitor with a resolution of 2560 by 1440 pixels.

In the first part of the study, we showed participants the two dashboards, one at a time. To alleviate a potential priming bias, we alternated the prototype presented first in the study.
The study started by introducing the data to the participants. While we used the original brain network data with the expert group, we changed the context for our lay participants. Our goal was to make the data more relatable so that no knowledge barrier might confuse the lay participants, and they could entirely focus on the visualizations. We replaced the medical terminology with more relatable terms. Specifically, the participants were told that the data about brain activity related to five different feelings (instead of frequency bands), which were supposed to be measured at the points showing the ROIs in the prototype. After the study, we told the participants about the EEG context, clarifying that the terminology was used to simplify and focus the task. 
In the next step, we introduced the first dashboard and explained its different parts, allowing questions during this demonstration. Afterward, the participants were asked to complete four tasks while thinking aloud.

We based the tasks on the findings from the earlier requirement phase (see \autoref{sec:requirement}). These showed that the domain experts might focus on the whole brain and specific parts of the brain and how these parts interact with each other. This is true for metrics on a nodal and a global level. Hence, for the first two tasks (task 1 and task 2), we asked the participants to consider all 72 ROIs (the whole brain) and only a subset for the two subsequent tasks (task 3 and task 4). We also found that the domain experts are especially interested in analyzing metrics across frequency bands, which is reflected in task 1 and task 3, and connectivity between ROIs, which is reflected in task 2 and task 4.

The following paragraphs show the task specification, which we provided as a printout to the participants so that they could revisit it during the study. Additionally, we explain how the tasks are solved with each prototype. This explanation was not provided to the participants during the study. We relate each evaluation task to the high-level tasks identified in \autoref{sec:requirement}, as seen in parentheses.

\textbf{Task 1} (\textit{Overview, Locate, Compare}): ``Which of the five frequencies has the highest metric value?'' (domain experts) / ``Which of the five feelings has the highest metric value?'' (lay people)
\begin{itemize}
    \item Prototype A: Select the ring segment with the highest color saturation of all five frequency bands/feelings in the \textbf{ring view (2)}.
    \item Prototype B: Select the highest bar of all five frequency bands/feelings in the \textbf{bar view (2)}.
\end{itemize}
Follow-up question: ``Which region in this frequency has the highest metric value?'' (domain experts) / ``Which region in this feeling has the highest metric value?'' (lay people)
\begin{itemize}
    \item Prototype A: Identify the region's name with the highest color saturation in the previously found frequency band/feeling in the \textbf{ring view (2)}.
    \item Prototype B: Identify the name of the highest bar in the previously found frequency band/feeling in the \textbf{bar view (2)}.
\end{itemize}

\textbf{Task 2} (\textit{Overview, Locate, Compare}): 
``What is the strongest connection between these regions in each frequency band?'' (domain experts) / ``What is the strongest connection between these regions in each feeling?'' (lay people)
\begin{itemize}
    \item Prototype A: Identify the ROI-pair connected by the line with the highest color saturation in each frequency band/feeling in the \textbf{ring view (2)}.
    \item Prototype B: Identify the ROI-pair of the cell with the highest color saturation in each frequency band/feeling in the \textbf{heatmap view (3)}.
\end{itemize}

Before continuing with the next two tasks, we asked participants to create three subnetworks containing five, three, and five ROIs, respectively (see \autoref{fig:ringsub}).

\textbf{Task 3} (\textit{Cluster, Locate, Compare}): 
``In which frequency band does ROI 21 have the lowest metric value?'' (domain experts) / ``In which feeling does ROI 21 have the lowest metric value?'' (lay people)
\begin{itemize}
    \item Prototype A: Find the frequency band/feeling where ROI 21 has the lowest color saturation in \textbf{ring view (2)}.
    \item Prototype B: Find the frequency band/feeling where ROI 21 has the lowest bar in \textbf{bar view (2)}.
\end{itemize}

\textbf{Task 4} (\textit{Cluster, Locate, Compare}): 
``What is the strongest connection between these regions in each frequency band?'' (domain experts) / ``What is the strongest connection between these regions in each feeling?'' (lay people)
\begin{itemize}
    \item Prototype A: Identify the ROI-pair connected by the line with the highest color saturation in each frequency band/feeling in the \textbf{ring view (2)}.
    \item Prototype B: Identify the ROI-pair of the cell with the highest color saturation in each frequency band/feeling in the \textbf{heatmap view (3)}.
\end{itemize}

\subsection{Semi-structured Interview}
After completing the tasks, we conducted a semi-structured interview consisting of two parts. The first part aimed at general feedback (especially on effectiveness and efficiency) to both prototypes. In the second part, we asked the participants to compare the prototypes and let them rank one over the other for different use cases. Throughout the interview, participants had a screenshot of the two prototypes in front of them to avoid mixing them up. The complete interview outline can be found in the supplemental material.

\subsection{Questionnaire}
The study concluded with a questionnaire the participants completed without supervision right after the interview. The questionnaire included the system usability scale (SUS)~\cite{brooke1996} to assess the prototypes' usability, the BeauVis scale~\cite{he2023} to assess the aesthetic pleasure of the prototypes, an overall grading of the interfaces, and a demographic section. The questionnaire can be found in the supplemental material.

\section{Results}
\label{sec:results}
After the initial interviews about the low-fidelity prototypes with the domain experts (see \autoref{sec:lowfi}), the combined ring view was preferred over conventional bar charts because of its aesthetics and originality in the medical domain. However, visualization literature \cite{munzner2014,cleveland1984,mackinlay1986,heer2010} has shown that the visualization channels used to encode the data in prototype A (metric value encoded through color saturation) are usually less effective than the ones used in prototype B (metric value encoded as the length of the bar). Our evaluation results show that the initial preference changed: prototype B was the most effective and efficient choice. The results, therefore, concur with the visualization literature. The results are summarized in \autoref{tab:evaluation_results}.

In the following section, we present the results and findings of our user study. First, we discuss observations made during the task-based within-subject study, followed by a qualitative and quantitative analysis of the interview and questionnaire.

\begin{table}[t]
    \centering
    \caption{Summary of the evaluation results for high-fidelity prototypes A and B. The first block (a) shows the participants' prototype preference for different aspects. The average SUS score (0: worst; 100: best), BeauVis score (1: worst; 7: best), and grade (1: best, 5: worst) are shown in (b). Part (c) shows the average completion time per task for each prototype and participant group. The last part (d) shows the average score (1: very easy; 5: not easy at all) of how easy participants perceived solving the task with the respective prototype.}
    \bgroup
    \renewcommand{\arraystretch}{1.1}
    \begin{tabular}{ |c|c|c|c||c|c| }
      \hline
      \multirow{14}{*}{\rotatebox[origin=c]{90}{(a) Preference}} & & \multicolumn{2}{c||}{\textbf{Domain Experts}} & \multicolumn{2}{c|}{\textbf{Lay People}}\\
      & & \multicolumn{2}{c||}{(n=4)} & \multicolumn{2}{c|}{(n=12)} \\
      & & \textbf{Prot. A} & \textbf{Prot. B} & \textbf{Prot. A} & \textbf{Prot. B}\\
      \hline
      & \textbf{Overall} & 25\% & \cellcolor[HTML]{bdd7e7}\textbf{75}\% & 33.3\% & \cellcolor[HTML]{bdd7e7}\textbf{66.7}\%\\
      
      & \textbf{Task 1} & 50\% & 50\% & \cellcolor[HTML]{bdd7e7}\textbf{66.7\%} & 33.3\%\\
      
      & \textbf{Task 2} & 0\% & \cellcolor[HTML]{bdd7e7}\textbf{100\%} & 25\% & \cellcolor[HTML]{bdd7e7}\textbf{75\%}\\
      
      & \textbf{Task 3} & 25\% & \cellcolor[HTML]{bdd7e7}\textbf{75\%} & 16.7\% & \cellcolor[HTML]{bdd7e7}\textbf{83.3\%}\\
      
      & \textbf{Task 4} & 0\% & \cellcolor[HTML]{bdd7e7}\textbf{100\%} & 41.7\% & \cellcolor[HTML]{bdd7e7}\textbf{58.3\%}\\
      
      & \textbf{Overview} & 25\% & \cellcolor[HTML]{bdd7e7}\textbf{75\%} & \cellcolor[HTML]{bdd7e7}\textbf{58.3\%} & 41.7\%\\
      
      & \textbf{Value} & 25\% & \cellcolor[HTML]{bdd7e7}\textbf{75\%} & 25\% & \cellcolor[HTML]{bdd7e7}\textbf{75\%}\\
      
      & \textbf{Aesthetic} & \cellcolor[HTML]{bdd7e7}\textbf{100\%} & 0\% & \cellcolor[HTML]{bdd7e7}\textbf{91.7\%} & 8.3\%\\
      \hhline{======}
      \multirow{6}{*}{\rotatebox[origin=c]{90}{(b) Avg. Score}} & \textbf{SUS} & 73.8 & \cellcolor[HTML]{bdd7e7}\textbf{84.4} & 63.3 & \cellcolor[HTML]{bdd7e7}\textbf{85.4}\\
      & (0–100) & ($\sigma$=14.5) & \cellcolor[HTML]{bdd7e7}\textbf{($\sigma$=3.2)} & ($\sigma$=24.9) & \cellcolor[HTML]{bdd7e7}\textbf{($\sigma$=13.9)}\\
      
      & \textbf{BeauVis} & \cellcolor[HTML]{bdd7e7}\textbf{5.7} & 5.4 & \cellcolor[HTML]{bdd7e7}\textbf{5.5} & 5.3\\
      & (1–7) & \cellcolor[HTML]{bdd7e7}\textbf{($\sigma$=1)} & ($\sigma$=0.9) & \cellcolor[HTML]{bdd7e7}\textbf{($\sigma$=1.3)} & ($\sigma$=0.9)\\
      
      & \textbf{Grade} & 2.3 & 2.3 & 2.4  & \cellcolor[HTML]{bdd7e7}\textbf{1.8}\\
      & (1–5) & ($\sigma$=1.5) & ($\sigma$=0.5) & ($\sigma$=1) & \cellcolor[HTML]{bdd7e7}\textbf{($\sigma$=0.6)}\\
      \hhline{======}
      \multirow{8}{*}{\rotatebox[origin=c]{90}{(c) Avg. Time [sec]}} & \multirow{2}{*}{\textbf{Task 1}} & 98 & \cellcolor[HTML]{bdd7e7}\textbf{65} & 104 & \cellcolor[HTML]{bdd7e7}\textbf{58}\\
      & & ($\sigma$=69) & \cellcolor[HTML]{bdd7e7}\textbf{($\sigma$=31)} & ($\sigma$=58) & \cellcolor[HTML]{bdd7e7}\textbf{($\sigma$=40)}\\
      
      & \multirow{2}{*}{\textbf{Task 2}} & 217 & \cellcolor[HTML]{bdd7e7}\textbf{164} & 213 & \cellcolor[HTML]{bdd7e7}\textbf{208}\\
      & & ($\sigma$=80) & \cellcolor[HTML]{bdd7e7}\textbf{($\sigma$=41)} & ($\sigma$=76) & \cellcolor[HTML]{bdd7e7}\textbf{($\sigma$=68)}\\
      
      & \multirow{2}{*}{\textbf{Task 3}} & 62 & \cellcolor[HTML]{bdd7e7}\textbf{26} & 53 & \cellcolor[HTML]{bdd7e7}\textbf{32}\\
      & & ($\sigma$=58) & \cellcolor[HTML]{bdd7e7}\textbf{($\sigma$=13)} & ($\sigma$=34) & \cellcolor[HTML]{bdd7e7}\textbf{($\sigma$=26)}\\

      & \multirow{2}{*}{\textbf{Task 4}} & 150 & \cellcolor[HTML]{bdd7e7}\textbf{77} & 80 & \cellcolor[HTML]{bdd7e7}\textbf{75}\\
      & & ($\sigma$=132) & \cellcolor[HTML]{bdd7e7}\textbf{($\sigma$=34)} & ($\sigma$=14) & \cellcolor[HTML]{bdd7e7}\textbf{($\sigma$=26)}\\
      \hhline{======}
      \multirow{8}{*}{\rotatebox[origin=c]{90}{(d) Avg. Easiness}} & \textbf{Task 1} & \cellcolor[HTML]{bdd7e7}\textbf{2} & 2.5 & \cellcolor[HTML]{bdd7e7}\textbf{1.3} & 1.8\\
      & (1–5) & \cellcolor[HTML]{bdd7e7}\textbf{($\sigma$=1.4)} & ($\sigma$=1.3) & \cellcolor[HTML]{bdd7e7}\textbf{($\sigma$=0.5)} & ($\sigma$=1)\\
      
      & \textbf{Task 2} & 2.3 & \cellcolor[HTML]{bdd7e7}\textbf{1.3} & 2.3 & \cellcolor[HTML]{bdd7e7}\textbf{1.8}\\
      & (1–5) & ($\sigma$=1) & \cellcolor[HTML]{bdd7e7}\textbf{($\sigma$=0.5)} & ($\sigma$=1.2) & \cellcolor[HTML]{bdd7e7}\textbf{($\sigma$=0.9)}\\
      
      & \textbf{Task 3} & 2.5 & \cellcolor[HTML]{bdd7e7}\textbf{1.8} & 2.1 & \cellcolor[HTML]{bdd7e7}\textbf{1.5}\\
      & (1–5) & ($\sigma$=1.7) & \cellcolor[HTML]{bdd7e7}\textbf{($\sigma$=0.5)} & ($\sigma$=1) & \cellcolor[HTML]{bdd7e7}\textbf{($\sigma$=0.7)}\\

      & \textbf{Task 4} & 1.5 & \cellcolor[HTML]{bdd7e7}\textbf{1.3} & 1.4 & \cellcolor[HTML]{bdd7e7}\textbf{1.3}\\
      & (1–5) & ($\sigma$=0.6) & \cellcolor[HTML]{bdd7e7}\textbf{($\sigma$=0.5)} & ($\sigma$=0.9) & \cellcolor[HTML]{bdd7e7}\textbf{($\sigma$=0.7)}\\
      \hline
    \end{tabular}
    \egroup
    \label{tab:evaluation_results}
\end{table}

\subsection{Observations}
We conducted one-on-one sessions, each taking about 1 hour (average time: 60 minutes for domain experts, 58 minutes for lay people). The participants were asked to think aloud while solving the tasks. During the study, they could ask questions if something was unclear, and we observed the participants' interaction with the prototypes. Additionally, we measured the time participants needed to complete the tasks and recorded the number of wrong answers (error count).

On average, solving the tasks with prototype A took domain experts longer than with prototype B. While they needed between 5 and 13.4 minutes ($\mu$ = 8.8 min; $\sigma$ = 4.4 min) with prototype A, they spent between 4.3 and 6.9 minutes ($\mu$ = 5.5 min; $\sigma$ = 1.3 min) with prototype B. These results show that domain experts could solve the tasks more efficiently using prototype B. 
The lay participants, who spent between 3.9 and 11.4 minutes ($\mu$ = 7.5 min; $\sigma$ = 2 min) with prototype A and between 3.5 and 9.4 minutes ($\mu$ = 6.2 min; $\sigma$ = 1.9 min) with prototype B, were also more efficient using prototype B. Except for task 2, the individual completion times confirm this result (see \autoref{tab:evaluation_results}c).
\\\\
\textbf{Task 1}. Domain experts needed about two-thirds (66.3\%), while lay people only needed about half (55.8\%) of the time used for identifying the ring segment with the highest color saturation in the \textit{ring view} (lay: 104s; experts: 98s) to find the overall highest bar in the \textit{bar chart view} (lay: 58s; experts: 65s). 
Most participants filtered for the top [X] percentage of metric values before identifying the highest value visually. While lay participants did not give any wrong answers for the tasks using prototype B, four initially gave a wrong answer when working with prototype A. Domain experts made one error using prototype A. In contrast, they did not answer incorrectly when working with prototype B.
\\\\
\textbf{Task 2}. Finding the ROI-pair with the highest connectivity value in each frequency in the \textit{heatmap view} in prototype B (164s) took the domain experts only about three-fourths (75.6\%) of the time compared to the \textit{ring view} in prototype A (217s). This result might be influenced by the domain experts' familiarity with heatmaps since the time needed by the lay people was nearly the same for prototype B (208s) and prototype A (213s).
Concerning the error rate, the domain experts gave one wrong answer per prototype for this task, while lay people made two wrong conclusions using prototype A and gave three wrong answers using prototype B.
\\\\
\textbf{Task 3}. Solving this task with prototype B took domain experts only about 42\% of the time needed with prototype A. Another speed-up was observed with lay participants: they solved the task with prototype B in 60.4\% of the time needed with prototype A. Even though the participants had to compare bars on top of each other rather than being aligned on a common x-axis, they still needed less time than comparing the color saturation of the ring segments. While the solution of two domain experts was initially wrong using prototype A, there were no errors using prototype B. All lay participants solved the tasks correctly with prototype B, while only one lay person gave a wrong answer with prototype A.
\\\\
\textbf{Task 4}. The last task was similar to task 2, with the only difference being that the task had to be solved for a subset of 13 ROIs. Domain experts were almost twice as fast with prototype B (77s) than with prototype A (150s). The difference in completion times for lay participants is not as significant as for the domain experts: lay people only needed five seconds longer with prototype A (80s) compared to prototype B (75s). While domain experts did not give any wrong answers using both prototypes, only one lay participant gave a wrong answer using prototype A.
\\\\
We also asked participants to rate how easy it was to solve each task with the prototypes (1: very easy; 5: not easy at all). While participants found solving most tasks easier using prototype B (see \autoref{tab:evaluation_results}d), therefore confirming the results above, task 1 was considered easier using prototype A. The participants mentioned this mainly because they could see the distribution of the metric values directly in the \textit{metric histogram view}, not the \textit{ring view}.

\subsection{Interview}
We recorded all interviews and took notes during the interview. For the analysis, we extracted themes from the gathered knowledge and clustered them for each participant group individually. In the following subsections, we present our interview findings.

\subsubsection{Interviews with Domain Experts} 
While the layered doughnut chart (paper prototype 3) was favored by the domain experts in the low-fidelity interview study, three out of the four experts (E$_{1,3,4}$) in the final study preferred prototype B \textit{(bar charts and heatmap)} in terms of effectiveness and efficiency. They found interface B especially helpful for quickly identifying connectivity patterns in the heatmap (E$_{1,4}$). This might be related to their familiarity and experience with this type of visualization. Another participant (E$_{3}$) mentioned that he enjoyed having separate bar charts for the metric in each frequency as this allowed him to get a global overview of all ROIs. He also noted that the space was used more efficiently than in prototype A \textit{(layered doughnut chart)}, resulting in less whitespace overall. However, two experts (E$_{2,4}$) found it hard to compare bars across frequencies since the corresponding bar charts were on top of each other. One expert (E$_{3}$) also mentioned his concerns about prototype B not scaling well for more than 72 regions since the screen real estate is limited and comparing bars might get more difficult. Suggestions for improvement that were mentioned include using a dynamic or logarithmic color scale to facilitate the differentiation of the color saturation in the \textit{heatmap view} (E$_{3,4}$), encoding metric values additionally through color saturation in the \textit{bar view} (E$_{4}$), and implementing a feature to save and pre-load subnetworks (E$_{3}$). While three experts (E$_{2,3,4}$) saw the potential for the current version of prototype B to facilitate data analysis in their daily work, one expert (E$_{1}$) reported he would need more customization features before he would use it daily.

Only one of the four participants (E$_{2}$) preferred prototype A. He liked the overall presentation of the connections and frequencies in the aggregated \textit{ring view} and the filter option only showing connections of interest. Even though he initially found prototype B easier to use, he supports his choice for prototype A by saying, ``[prototype B] being easier today does not mean that it is always easier – prototype A gets better over time.''
The main advantage of prototype A that was mentioned (E$_{1,2,4}$) is the compact \textit{ring view} that allows for a good overview and presentation of the metric and connectivity values across all frequencies. E$_{1}$ said that ``[prototype] B is better for analyzing the data; [prototype] A is better for presenting the data''. E$_{3}$ commented that if he ``had both tools, [he] would only use [prototype] B. [He is] not the biggest fan of A as it is not so easy to interpret''. 

These results underline our findings in the other two parts of the study, which state that prototype A is still preferred in terms of aesthetics. Still, prototype B is considered more effective and efficient for data analysis. When asked about potential use cases for prototype A over prototype B, our participants mentioned situations in which they must present the data (e.g., team meetings, research papers) and want to attract attention. The \textit{histogram views} to filter data attributes were also well received by three experts (E$_{1,3,4}$) who suggested including them in prototype B.
Two participants (E$_{1,4}$) mentioned that it is difficult to see exact values in prototype A since they are mainly encoded through color saturation. Related to that, one participant (E$_{3}$) found it particularly ``difficult to compare the color saturation in the small ring segments''.
While all experts saw a potential benefit of prototype A for their daily work (especially for data presentation), most of them (E$_{1,3,4}$) still preferred prototype B for analyzing the data as it is ``intuitive and easier to use than the other prototype''.

\subsubsection{Interviews with Lay People} 
The lay participants' responses mostly aligned with the domain experts' opinions. The results show that eight of them (P$_{2, 4, 5, 6, 9, 10, 11, 12}$) preferred prototype B \textit{(bar charts and heatmap)}. They based their decision on several reasons, one being that it is easier for them, especially in the \textit{bar view}, to see data values more accurately than in prototype A \textit{(layered doughnut chart)} (P$_{1, 3, 4, 6, 8, 10, 11, 12}$). More concretely, four participants (P$_{3, 5, 7, 9}$) explicitly pointed out that they preferred the bars over the encoding with color. The encoding of the connectivity values in the \textit{heatmap view} through a heatmap was received well (P$_{7, 9, 10}$). They point out that the heatmap conveys more information visually than the \textit{ring view} in prototype A (P$_{4, 6, 8}$) and that the color saturation is easier to distinguish in a horizontal and vertical grid than in a circular layout (P$_{2, 10, 12}$). Three participants (P$_{5, 6, 8}$) found prototype B more user-friendly and intuitive than the other prototype. The main weakness identified in prototype B was related to the scalability of the interface. Since the initial interface features all 72 ROIs, participants (P$_{1, 8, 9}$) found that there is an overload of visual elements on the screen when all regions are shown and $P_{1}$ added that comparing bars is difficult when there are so many of them. Therefore, the interface would not scale well as more ROIs are added, making the analysis more difficult (P$_{1, 3, 5}$). Since the connectivity matrices in our data set are all symmetric, half of the participants (P$_{2, 3, 7, 8, 10, 11}$) found it unnecessary to display the whole heatmap. This was an explicit design decision, as our requirement analysis showed that connectivity matrices in other data sets might not be symmetric. It is, however, a suggestion for improvement to allow the user only to display half of the heatmap.

Even though most participants preferred prototype B overall, almost half of them (P$_{1, 3, 4, 5, 8}$) enjoyed the clean and compact representation of the data in prototype A. P$_{1, 5, 6, 8}$ liked the aggregation of the data attributes (metric and connectivity) in a single \textit{ring view}. Additionally, all participants pointed out that the \textit{metric histogram view}, as well as the \textit{connectivity histogram view}, are very helpful, primarily since they can be used as a filter for the \textit{ring view}.
The main weaknesses identified in prototype A are that the usability is worse than in the other prototype (P$_{5, 6, 8, 9, 10}$) and that training (P$_{6, 8, 9}$) is needed before the tool can be used efficiently. P$_{5, 7, 9, 10, 11}$ mentioned that the \textit{ring view} and its ring segments are too small for the displayed data, thus negatively influencing the tool's usability. Participants also had a hard time differentiating color saturation in the \textit{ring view} for both data features: metric (P$_{2, 5, 7, 8, 10}$) and connectivity (P$_{8, 9, 10}$). This was underlined by P$_{1}$, who said that ``[data feature] values are easier estimated through bars than color saturation''; P$_{3}$ found that ``hovering for comparing segments with higher color saturation is tedious''. All participants agreed that they find prototype A more aesthetic. One of the participants (P$_{6}$) commented, ``I'm just not a big fan of the circle. However, if I had to visualize results and print them in a research paper, then I would go with [prototype] A''. A few lay participants (P$_{2, 4, 9}$) also pointed out that in both prototypes the \textit{brain view} did not add any value for them since the brain representation is for users with domain knowledge who can interpret the data.

\subsubsection{Suggestions for Improvement}
Both participant groups shared suggestions to improve the prototypes after completing a task or in the interview afterward.
\\\\
\noindent \textbf{Prototype A}. Two participants (P$_{6}$, E$_{1}$) were interested in seeing the connectivity values in prototype A, so they suggested displaying the value when the user hovers over a connection. E$_{1}$ further mentioned that he would be interested in simultaneously seeing connections from multiple frequencies. However, displaying too many connections in the \textit{ring view} could result in an unreadable mesh of lines. E$_{3}$ suggested adding line thickness as a second encoding channel for the connectivity value to make strong connections even more noticeable to the analysts. Concerning data filtering, some participants (P$_{1, 2, 4, 9}$) wanted additional lower-level filter options (e.g., letting users input a custom percentage for the top percentages of values). One expert (E$_{3}$) mentioned that he had a hard time identifying bins that only contain a few elements in the \textit{connectivity histogram view} when there were bins containing high numbers of elements. He suggested rescaling the histogram when the user selects a range to alleviate this problem.

\noindent \textbf{Prototype B}. Both histogram views (connectivity and metric) benefitted all participants in prototype A, making it evident that including them in prototype B could improve it further. This is also indicated by participants' suggestions to add more filter options (P$_{1, 2}$) and allow users to choose custom percentages for the top [X] values (P$_{2, 4}$). Another feature request would allow users to sort bar charts in ascending/descending order (P$_{2, 6}$) to identify maxima and minima quickly. Two participants (P$_{6}$, E$_{4}$) suggested adding color saturation as a second encoding channel for the metric value in the bar chart to help users identify higher values more easily. A dynamic color scheme updating whenever a filter or subnetwork selection is applied was suggested (E$_{1,4}$) for the \textit{heatmap view} to distinguish data points easily. Loading pre-defined subnetworks and saving custom subnetwork selections could save the user time during the configuration (E$_{3}$).

\subsection{Questionnaire}
The quantitative analysis results mirrored the qualitative analysis's findings. The results of the questionnaire (see \autoref{tab:evaluation_results}b) show that prototype B has a higher SUS score for domain experts (84.4) as well as lay people (85.4) than prototype A (experts: 73.8; lay people: 63.3). This almost corresponds to ``excellent'' usability for prototype B, whereas the usability of prototype A is rated between ``ok'' and ``good''~\cite{bangor2008}. We used the BeauVis scale~\cite{he2023} in its recommended version: a 7-point Likert scale with five items (enjoyable, likable, pleasing, nice, and appealing) from ``strongly disagree'' (1) to ``strongly agree'' (7). Even though the usability score of prototype B is higher, the BeauVis rating of prototype B (experts: 5.4; lay people: 5.3) is close to the rating of prototype A (experts: 5.7; lay people 5.5). This shows that our effort for a clear and accessible design paid off by also getting high results in terms of aesthetics for prototype B. However, we did not reach the same level as prototype A, which is still better, and the result illustrates the participants' aesthetic preference for prototype A, as indicated clearly in the interviews.

We also asked participants to grade each prototype from 1 (best) to 5 (worst). While the average grade of domain experts is the same for both prototypes (2.3), lay people assigned better grades to prototype B (1.8) than to prototype A (2.4).

\section{Discussion}
\label{sec:discussion}
Quantitative and qualitative results show that we successfully developed effective and efficient prototypes for analyzing multi-frequency network data. The visual analysis tools facilitate the visual exploration of the data. Here we address important aspects of the design and evaluation process.

\noindent \textbf{The importance of high-fidelity prototyping.} The evaluation with domain experts in the early low-fidelity prototyping stages showed that their preference for a tool design was affected by its aesthetics and novelty. Another design proposal was less preferred even though it was considered more optimal based on visualization theory. However, the overall preference shifted once we implemented the low-fidelity designs as high-fidelity prototypes and allowed users to interact with the systems. This preference shift underlines the importance of low- and high-fidelity parallel prototyping. Visualization designers must create multiple prototypes in parallel to explore the design space. Furthermore, our findings in this study show that it is not enough to stop at the low-fidelity phase since preferences can change in further iterations when new factors (e.g., interactivity) come into play. Hence, more and better rapid prototyping tools are needed to bridge this gap between low- and high-fidelity prototyping.

\noindent \textbf{There is a lack of design guidelines for affective intent.}
Designing visualization goals solely based on their task-based objectives (e.g., efficiently conveying facts or insights) might not lead to the optimal solution. Other factors (e.g., aesthetics, engagement), which can be used to achieve affective intent~\cite{leerobbins2023}, must also be considered in characterizing the goal at the early stages of the requirement analysis. While our results show that prototype B outperformed prototype A in terms of effectiveness and efficiency for analyzing the data, prototype A was considered well-suited, especially for presenting the data, as it was considered more engaging. While one participant explicitly said, ``[prototype] B is better for analyzing the data; [prototype] A is better for presenting the data,'' other participants articulated similar thoughts agreeing with this statement. We argue that this trade-off between performance and aesthetics influences the design space significantly. The latter are often not explicit goals of expert visualization tools. Our results demonstrate that acknowledging these subjective types of goals changes the design. However, our community lacks design guidelines for affective intent. Our work shows that there is plenty of room to incorporate aesthetic aspects into prototype B. We believe creating engaging prototypes without sacrificing half the performance is possible, as was the case comparing our two prototypes.

\noindent \textbf{Be aware of familiarity with visual representations.} Similar to aesthetics, familiarity with specific visual representations could influence users' subjective preferences. Dasgupta~\cite{dasgupta2018} argues that ``familiar visualizations and visual encodings within a domain are often in conflict with the optimality of visualization design, in terms of how well they support the data analysis or communication tasks.''. This was not reflected in our findings in the initial interview study during the low-fidelity prototyping phase. Domain experts preferred the circular prototype to the one with bar charts and a heatmap regardless of their familiarity with heatmaps since they are prevalent in network analysis. However, this might have been because participants prioritized aesthetics and novelty over familiarity. We believe familiarity can play an important role in subjective preference and should be considered carefully while designing and evaluating a visual analysis tool.

\noindent \textbf{Limitations.}
We aimed to qualitatively and quantitatively validate and compare the expert group's findings with participants from outside the medical domain. Our non-representative participant sample was, in the case of the lay group, not biased or primed by domain knowledge or constrained by familiar visualizations. Hence, they could focus primarily on usability and aesthetics rather than functionality. This approach strengthened our findings and validated the expert group's results.

As a result of this design study, we developed two high-fidelity prototypes that showed great potential to improve the effectiveness and efficiency of data analysis for medical experts. However, to evaluate these factors of the systems in production, more implementation iterations are needed to cover all aspects of the data. A deployed productive system was out of scope for this project and is considered future work.

\section{Conclusion and Future Work}
\label{sec:conclusion}
In this paper, we reported on a design study for a visual analysis tool that facilitates exploratory data analysis of multi-frequency network data. We conducted an in-depth requirement analysis and developed two high-fidelity prototypes with domain experts in an iterative design process. We evaluated the prototypes regarding effectiveness, efficiency, and subjective preference both qualitatively and quantitatively. The evaluation results with domain experts and lay people, both with little or no visualization experience, showed that subjective preference and objective task performance do not always align. Furthermore, subjective preference can change throughout the design process, underlining the importance of high-fidelity prototyping, where users interact with their own data for the first time. We also found that subjective factors (e.g., aesthetics, familiarity) are important when defining design goals for bespoke visual analysis tools. However, as a community, we lack the ability to design for affective intent.

In the future, we envision further investigating the relationship between different subjective factors that influence the visualization goal and, therefore, also the tool design. Furthermore, we want to improve the prototypes based on the user study feedback and develop a consolidated version that can be deployed to production.

\section*{Acknowledgments}
We thank the domain experts and our study participants for their collaboration, support, and time. This research was supported in parts by the CHIST-ERA IVAN project (Award-ID: 20CH21\_174081) and by the Vienna Science and Technology Fund (WWTF) [10.47379/ICT20065].

\bibliographystyle{IEEEtran}
\bibliography{references}

\end{document}